\documentstyle[preprint,floats,aps,psfig]{revtex}

\begin{document}
\draft 
\thispagestyle{empty}
\firstfigfalse \firsttabfalse


\title{The reconstruction of Ni and Rh (001) surfaces \\ upon Carbon,
Nitrogen, or Oxygen adsorption}

\author{Dario Alf\`e,\cite{dario,email} Stefano de Gironcoli,\cite{email}
and Stefano Baroni\cite{email}} 

\address{SISSA -- Scuola Internazionale Superiore di Studi Avanzati
and \\
INFM -- Istituto Nazionale di Fisica della Materia \\
via Beirut 2-4, I-34014 Trieste, Italy}

\date{\today}
\maketitle

\begin{abstract}
Nickel and Rhodium (001) surfaces display a similar---as from
STM images---{\it clock reconstruction} when half a monolayer of
C/Ni, N/Ni or O/Rh is adsorbed; no reconstruction is observed instead
for O/Ni. Adsorbate atoms sit at the
center of the {\it black} squares of a chess-board, $c(2\times 2)$,
pattern and {\it two different} reconstructions are actually
compatible with the observed STM images---showing a $(2\times 2)p4g$
pattern---according to whether a rotation of the {\it black} or {\it
white} squares occurs. We report on a first-principles study of
the structure of X/Ni(001) and X/Rh(001) surfaces (X=C,N,O) at half a
monolayer coverage, performed using density-functional theory. Our
findings are in agreement with all available experimental information
and shed new light on the mechanisms responsible for the
reconstructions. We show that the same substrate may display different
reconstructions---or no reconstruction---upon adsorption of different
atomic species, depending on the relative importance of the chemical
and steric factors which determine the reconstruction.


\end{abstract}



\section{introduction}

Transition metal surfaces are attracting a wide scientific and
technological interest, especially because of their capability of
reducing the activation barrier of many important chemical
reactions. One particularly relevant example is the reaction 2CO + 2NO
$\rightarrow$ 2CO$_2$ + N$_2$, which eliminates the two poisonous CO
and NO gases from the exhaust gas of combustion engines. This reaction
is catalysed by many transition metal surfaces, among which rhodium
and platinum have been demonstrated to be the most efficient.

\begin{figure} \def\baselinestretch{1} \small
\centerline{ \psfig{figure=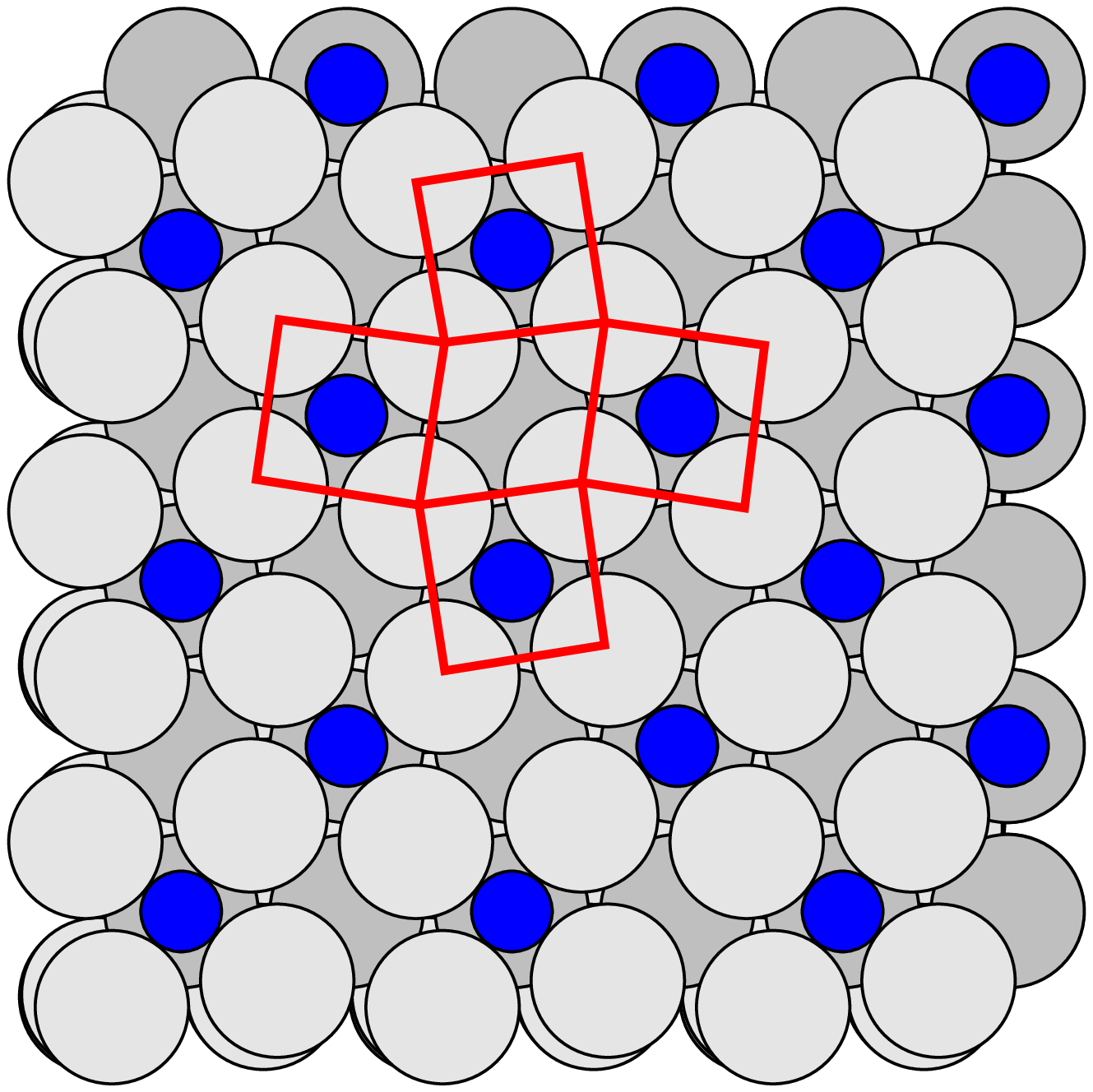,height=3.0in}
\psfig{figure=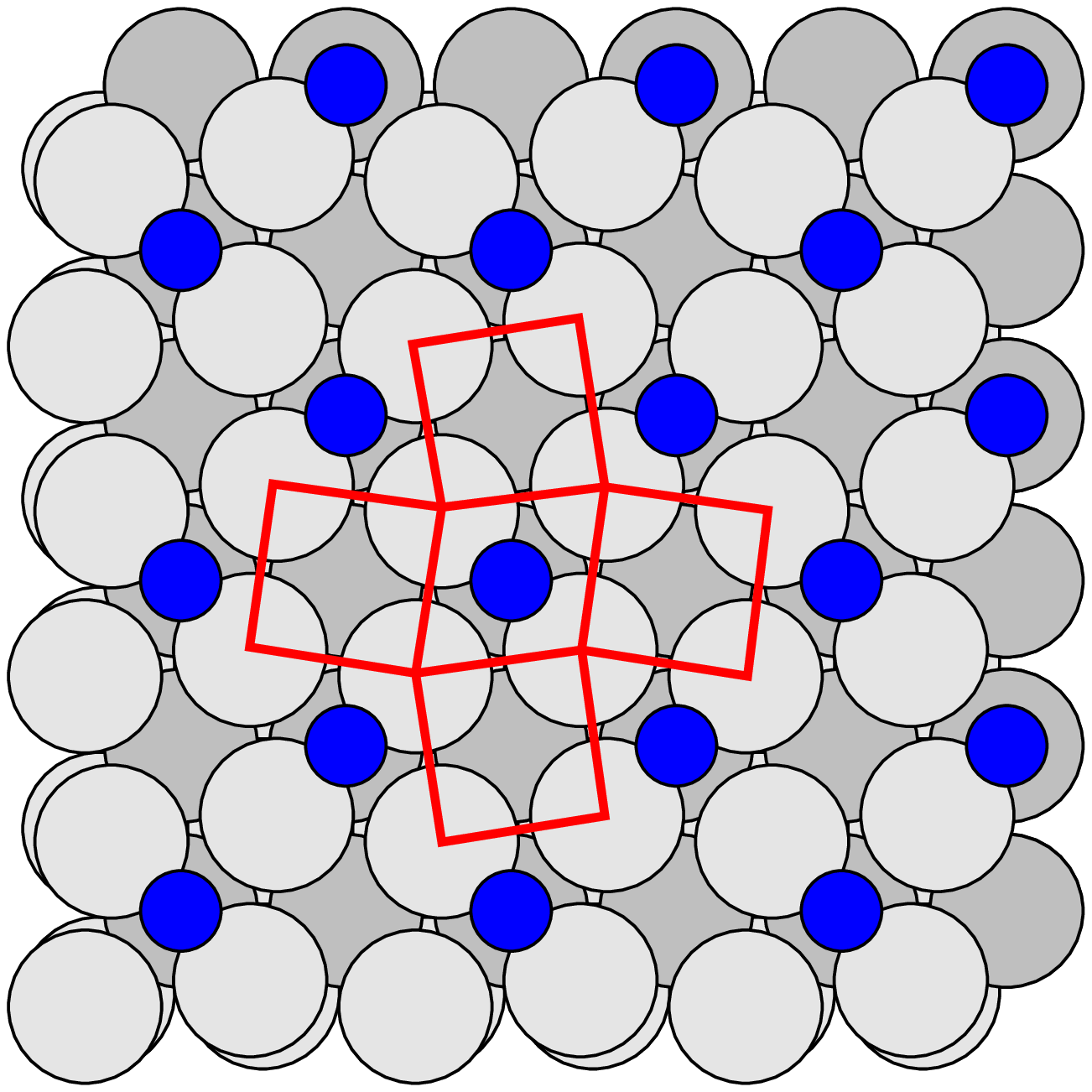,height=3.0in} } \vskip 8pt \caption{ {Two
possible {\it clock} reconstructions of the O/Rh(001) surface,
resulting in a $(2\times 2)p4g$ structure. In the {\it black}
reconstruction (a) the squares with an O atom in the middle rotate,
while the empty ones distort to rhombi; in the {\it white}
reconstruction (b), the opposite occurs.}}
\label{ricostruzione} \end{figure}

Adsorption of atomic or molecular species on surfaces modifies their
electronic and structural properties, thus affecting their catalytic
properties. Recently, the (001) surfaces of rhodium and nickel have
been studied with respect to their peculiar reconstructions upon
oxygen, carbon and nitrogen adsorption
\cite{Wat78,oed88,Mercer,noi,Daum86,onuferko79,klink93,leibsle93,wenzel,kilcoyne}.
Oxygen adsorption on Rh (001) is known to be dissociative and to
saturate at half a monolayer, independently of the adsorption
temperature. At this coverage a $(2\times 2)$ reconstruction has been
observed by LEED \cite{oed88} and confirmed by STM \cite{Mercer}. The
oxygen atoms sit in the troughs formed by four first-layer rhodium
atoms and fill these sites in a c($2\times 2$) geometry. This
structure may be seen as a chess-board whose {\it black} squares are
occupied by oxygen atoms, while the {\it white} ones are empty. Within
this picture, the reconstruction observed in Ref. \cite{Mercer} has
been described as a rotation of the {\it black} squares, resulting in
a ($2 \times 2$)p4g symmetry (see Fig. \ref{ricostruzione}a). This
distortion preserves the shape of the {\it black} squares, while the
{\it white} ones become rhomboid. A similar behavior is observed for
nitrogen and carbon adsorbed on the (001) surface of nickel, where the
rotation angle of the squares is much larger and the {\it clock}
reconstruction more evident %
\cite{Daum86,onuferko79,klink93,leibsle93,wenzel,kilcoyne}.

\begin{figure} \def\baselinestretch{1}
\centerline{\psfig{figure=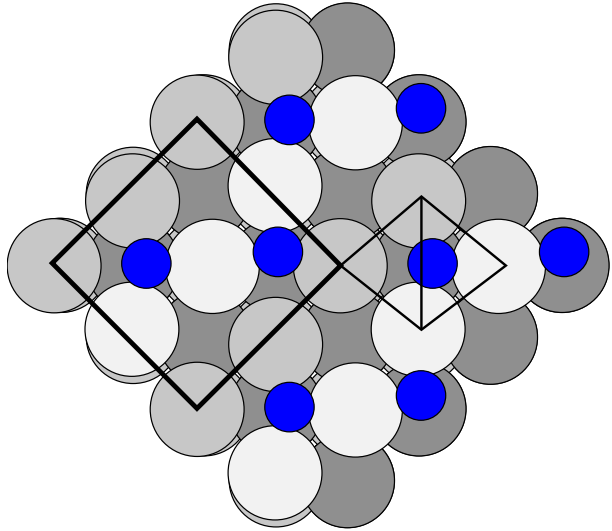,height=3in}} \vskip 8pt
\caption{\small Equilibrium structure of O/Rh(001) as obtained by our
simulated-annealing procedure. The thick line indicates the unit cell.
The thin line indicates the rhombus and its shorter diagonal. The
oxygen atoms are alternatively shifted orthogonally with respect to
the shorter diagonal. The atoms of the first surface layer depicted
with a brighter tone lean $\approx 0.08 $ \AA\ outwards with respect
to the others.}\label{nuovastruttura} \end{figure}

In our previous work on Rh(001)/O \cite{noi} we pointed out that a
different substrate reconstruction is actually compatible with LEED
and STM data. Using only these two experimental techniques it is not
possible to distinguish between the reconstruction described above and
the one which results instead from the rotation of the {\it white}
squares (Fig. \ref{ricostruzione}b). Our calculations actually
indicated that it is the {\it white} squares which rotate rather than
the {\it black} ones. Moreover, we predicted a reconstruction pattern
with a slightly different symmetry from what appears in the STM
pictures. We found that the ad-atoms get off the center of the
distorted squares, thus resulting in an {\it asymmetric clock}
reconstruction (Fig. \ref{nuovastruttura}). The oxygen atoms can
occupy two equivalent low-symmetry sites separated by a low energy
barrier, of the order of the room thermal energy. This would give rise
to an order-disorder transition, and we suggested that what is
actually observed in the STM pictures is the average position of the
oxygen atoms jumping back and forth between the two equilibrium
positions, in the high-temperature disordered phase.

The reconstruction of the (001) surface of nickel upon C and N
adsorption is experimentally much better characterized since also
LEED-IV \cite{onuferko79} and SEXAFS \cite{wenzel} data exist. In this
case it is widely accepted that the reconstruction is of the {\it
black} type, i.e. it is the filled squares which rotate
(Fig. \ref{ricostruzione}a). No surface reconstruction is induced by
oxygen adsorption on Ni(001) \cite{oed}, while as far as we know no
experimental data exist for carbon or nitrogen adsorbed on the Rh(001)
surface.

In this paper we present an {\it ab-initio} study of the adsorption of
carbon, oxygen, and nitrogen on the (001) surfaces of rhodium and
nickel. On rhodium, we find that {\it neither} nitrogen, {\it nor}
carbon, induce any reconstruction of the surface. On nickel, in
agreement with the experiments, we find that {\it both} carbon and
nitrogen induce a {\it clock} reconstruction, while oxygen adsorbs
with no induced reconstruction. We discuss the
interplay of the chemical and the steric effects which is at the root
of these behaviors.

The paper is organized as follows. In section \ref{method} we describe
our theoretical method, including some tests on the properties of the
rhodium and nickel bulk metals, and on those of the CO, O$_2$, NO and
N$_2$ molecules. In section \ref{results} we present our
results. Section \ref{discussion} contains the discussion and our
conclusions.

\section {Method}\label{method}

Our calculations are based on density functional theory within the
local-density approximation (LDA) \cite{DFT-LDA}, using Ceperley-Alder
exchange-correlation (XC) energies \cite{CeAl}. The one-particle
Kohn-Sham equations are solved self-consistently using plane-wave
basis sets in a {\it ultra-soft} (US) pseudo-potential scheme
\cite{Va90}. The rhodium, oxygen, and carbon pseudo-potentials are the
same as in Refs. \cite{StBa,alba}. For nitrogen and nickel we have
constructed new pseudo-potentials. In the case of Ni, we started from
the $3d^84s^24p^0$ reference configuration treating the $d$ channel in
the US scheme, while the $s$ and the $p$ channels were assumed to be
norm conserving. We chose the $p$ component of the pseudo-potential as
local part (this choice avoids the appearance of {\it ghost}
states). For N we started from the $2s^22p^3$ reference configuration
treating both the $s$ and the $p$ channels in the US scheme. Plane
waves up to maximum kinetic energy of 30 Ry are included in the basis
set. Brillouin-zone integrations have been performed using the
Gaussian-smearing \cite{MePa} special-point \cite{MoPa} technique. For
the calculations in the bulk, we have used a smearing function of
order 1 with a width $\sigma = 0.03$~Ry and a set of special points
equivalent to 10 {\bf k}-points in the irreducible wedge of the
Brillouin zone (IBZ). The convergence of our results with respect to
the size of the basis and special-point sets has been carefully
checked.

The isolated surfaces are modeled by a periodically repeated
super-cell. We have used the same super-cell for both the clean and
the covered surfaces. For the clean surfaces we have used $5$ atomic
layers plus a vacuum region corresponding to 6 layers. For the covered
surfaces, the $5$ Rh/Ni layers are completed by one layer of C, O, or
N atoms on each side of the slab; in this case the vacuum region is
correspondingly reduced to $\approx 5$ atomic layers. We have used a
Gaussian-smearing function of width $\sigma=0.03 \rm ~Ry$ and a
$(12,12,2)$ Monkhorst-Pack mesh \cite{MoPa} resulting in $21$ special
{\bf k}-points in the $1\times 1$ surface IBZ. Convergence tests
performed with a value of $\sigma$ twice as small and a
correspondingly finer mesh of special points resulted in no
significant changes in total energies and equilibrium geometries.

\subsection{Case tests on bulk metals and molecules}\label{tests}

In order to test the quality of our pseudo-potentials, we have done
some tests on rhodium and nickel bulks, as well as on CO, NO, O$_2$,
and N$_2$ molecules. To calculate the properties of the molecules the
latter have been put in a large cubic cell, with side L = 10 a.u, and
periodic boundary conditions. We have checked that our calculated
molecular properties were well converged with respect to the size of
the cell. The results of these tests are summarized in
Tab. \ref{bulks} and Tab. \ref{mol}.

\begin{table}[h!]
\def\baselinestretch{1}
\vskip 10pt
\caption{Theoretical structural parameters of FCC rhodium and nickel
obtained with a fit to Murnaghan equation of state of the calculated
energies as a function of the volume. Experimental data \protect
\cite{web} are reported for comparison.}
\begin{tabular}{lccc}
 & & Lattice constant & Bulk modulus \\
 &  &  (\AA)  & (Mbar) \\
\hline
Rh & Present work & 3.81 &  3.17 \\
 & Experiment & 3.80 &  2.69 \\
 & & & \\
Ni & Present work & 3.44 & 2.4 \\
 & Experiment & 3.52 & 1.88 \\
\end{tabular} \label{bulks}


\vskip 20pt

\caption{ Theoretical equilibrium distances and fundamental vibrational
frequencies. Experimental data \protect \cite{bransden} are
reported for comparison.}
\begin{tabular}{lccc}
 & & Equilibrium distance & Vibrational frequency \\
 &  &  (\AA)  & (cm$^{-1}$) \\
\hline
CO & Present work & 1.14 &  2200 \\
 & Experiment & 1.13 &  2170 \\
 & & & \\
NO & Present work & 1.16 & 2000 \\
 & Experiment & 1.15 & 1904 \\
 & & & \\
O$_2$ & Present work & 1.23 & 1650 \\
 & Experiment & 1.21 & 1580 \\
 & & & \\
N$_2$ & Present work & 1.11 & 2450 \\
 & Experiment & 1.09 & 2360 \\
\end{tabular}
\label{mol}
\end{table}

\begin{figure}
\def\baselinestretch{1}
\centerline{\psfig{figure=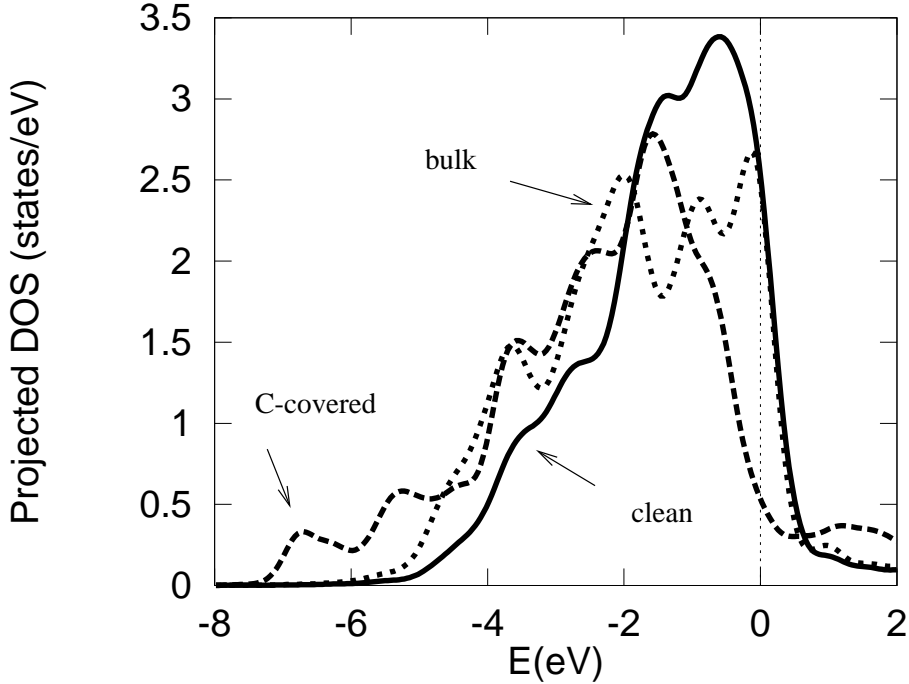,height=3.6in}}
\vskip 8pt
\caption{ 
$d$-projected density of states of nickel for the bulk, the clean surface,
and the carbon-covered surface.}
\label{nikeldos}
\end{figure}

Nickel and rhodium have the FCC crystal structure. To calculate the
equilibrium lattice constant $a_0$ and the bulk modulus $B_0$, we have
fitted the calculated energies as a function of the unit cell volume
to a Murnaghan's equation of state. Nickel is a magnetic metal with a
magnetic moment of 0.59 $\mu_B$/atom \cite{sazaki}. An explicit
account of the spin polarization within the local spin density
approximation did not result in any meaningful changes of the
calculated structural properties. The magnetism is expected to be more
important at the surface, because the surface density of states (SDOS)
is narrower than the bulk one. However, at least for the carbon- and
the nitrogen-covered surfaces, we have found that the Ni-$d$-SDOS is
shifted towards lower energies with respect to the bulk $d$ partial
density of states (see Fig. \ref{nikeldos}) thus resulting in a lower
density of states at the Fermi level. We argue that magnetic effects
should not be very important and they have been neglected altogether
in the present investigation.

\section{results}\label{results}

In order to find the stable surface structure of the C-, O-, and
N-covered Rh(001) and Ni(001) surfaces, we have performed simulated
annealing runs of the surfaces at half a monolayer of coverage, using
Born-Oppenheimer {\it ab-initio} molecular dynamics (AIMD). In our
implementation of AIMD the exact electronic ground state (within a
self-consistent threshold) is calculated at each time step using a
band-by-band conjugate gradient algorithm \cite{cg} and Broyden-like
\cite{broyden} mixing of the potentials. The forces are calculated
fully quantum mechanically, and the ionic equation of motion is
integrated using the Verlet algorithm \cite{Verlet}. We have used a
time step of 50 a.u ($\approx 1.2$ fs). After an equilibration period
of about 0.5 ps at a temperature of $\approx 800$ K we have slowly
cooled down the systems. Finally, when the ionic minimum structure is
reached, an accurate structure optimization is performed, by allowing
all the atoms in the slab to relax until the force acting on each of
them is smaller than $0.5\times 10^{-3} {\rm Ry}/{\rm a_{0}} $. Our
results are summarized in Table \ref{tab:riassunto} and discussed in
detail in the following sections.

\begin{table} \def\baselinestretch{1} \caption{Summary of the
structural data for the six systems investigated in this work. For the
reconstruction type see Fig. \ref{ricostruzione}. $d_{01}$ is the
distance between the adatoms and the first metal layer; $\delta$ is
the amplitude of the in-plane displacement of the first-layer metal
atoms upon reconstruction; $\Delta$E is the energy difference between
the symmetric c($2 \times 2$) structure and the reconstructed
one. Units are \AA\ for lengths and eV/atom for energies}
\begin{tabular}{lccccccc} & \multicolumn{2}{c}{Reconstruction }&
\multicolumn{2}{c}{$\delta$} & \multicolumn{2}{c}{ $d_{01}$} &
$\Delta$E \\ & Expt & Theory & Expt & Theory & Expt & Theory & Theory
\\ \hline C:Ni & {\it black} & {\it black} & $0.55 \pm 0.20
\tablenotemark[1]$ & $ 0.46 $ & $0.1\pm 0.1 \tablenotemark[1]$ &
$0.17$ & $0.20$ \\ N:Ni & {\it black} & {\it black} & $ 0.55 \pm 0.20
\tablenotemark[1]$ & $ 0.41 $ & $0.1 \pm0.1\tablenotemark[1]$ & $0.10$
& $0.08$ \\ O:Ni & none & none & & & $0.77 \pm 0.04\tablenotemark[2]$
& $0.73$ & \\ & & & & & & & \\ C:Rh & & none & & & & $0.63$ \\ N:Rh &
& none & & & & $0.71$ \\ O:Rh & {\it white}\tablenotemark[3] & {\it
white} & $0.20\pm 0.07 \tablenotemark[3] \tablenotemark[4]$ & $0.21$ &
$1.05 \pm 0.05 \tablenotemark[3] ~0.6\pm 0.1 \tablenotemark[4]$ &
$0.98 - 1.06 \tablenotemark[5]$ & $0.03$ \\
\end{tabular}
\tablenotetext[1]{From Ref. \cite{kilcoyne} and references quoted therein}
\tablenotetext[2]{From Ref. \cite{oed}}
\tablenotetext[3]{From Ref. \cite{ELETTRA}}
\tablenotetext[4]{See Ref. \cite{shen}}
\tablenotetext[5]{The two values correspond to the two Rh rows made
inequivalent by the {\it asymmetric} clock reconstruction (see text
and Fig. \ref{nuovastruttura}).}
\label{tab:riassunto}
\end{table}

\subsection{Rhodium}

Results for the O/Rh(001) surface have been reported and extensively
discussed in our previous work \cite{noi}. For this system we have
found that oxygen induces an {\it asymmetric clock} reconstruction of
{\it white} type (Fig. \ref{nuovastruttura}), where the oxygen-filled
sites are deformed into rhombi, and the ad-atoms get off the center of
the rhombi. The asymmetric position of the oxygen ad-atoms results in
a buckling reconstruction of the first surface layer, and the rhodium
rows are alternatively shifted up and down by $\approx 0.08$ \AA.
Oxygen atoms tend to shorten the bonds with the neighboring rhodium
atoms, and this results in a deformation of the oxygen site which
becomes a rhombus. We argued that the driving mechanism for this
reconstruction is the re-bonding of the oxygen with the first layer
rhodium surface. This can be better understood by the inspection of
the local density of states (LDOS).

\begin{figure}
\def\baselinestretch{1}
\centerline{\psfig{figure=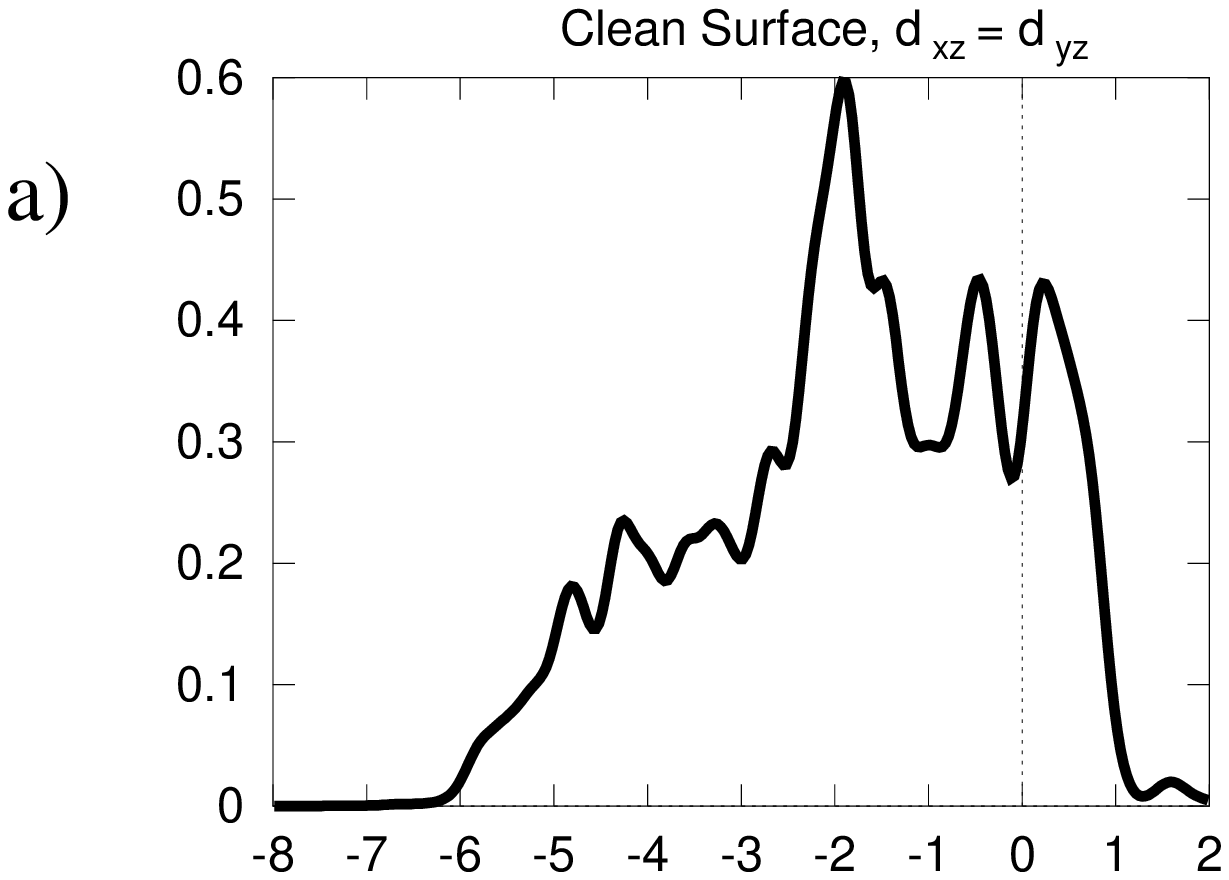,height=2in}
\psfig{figure=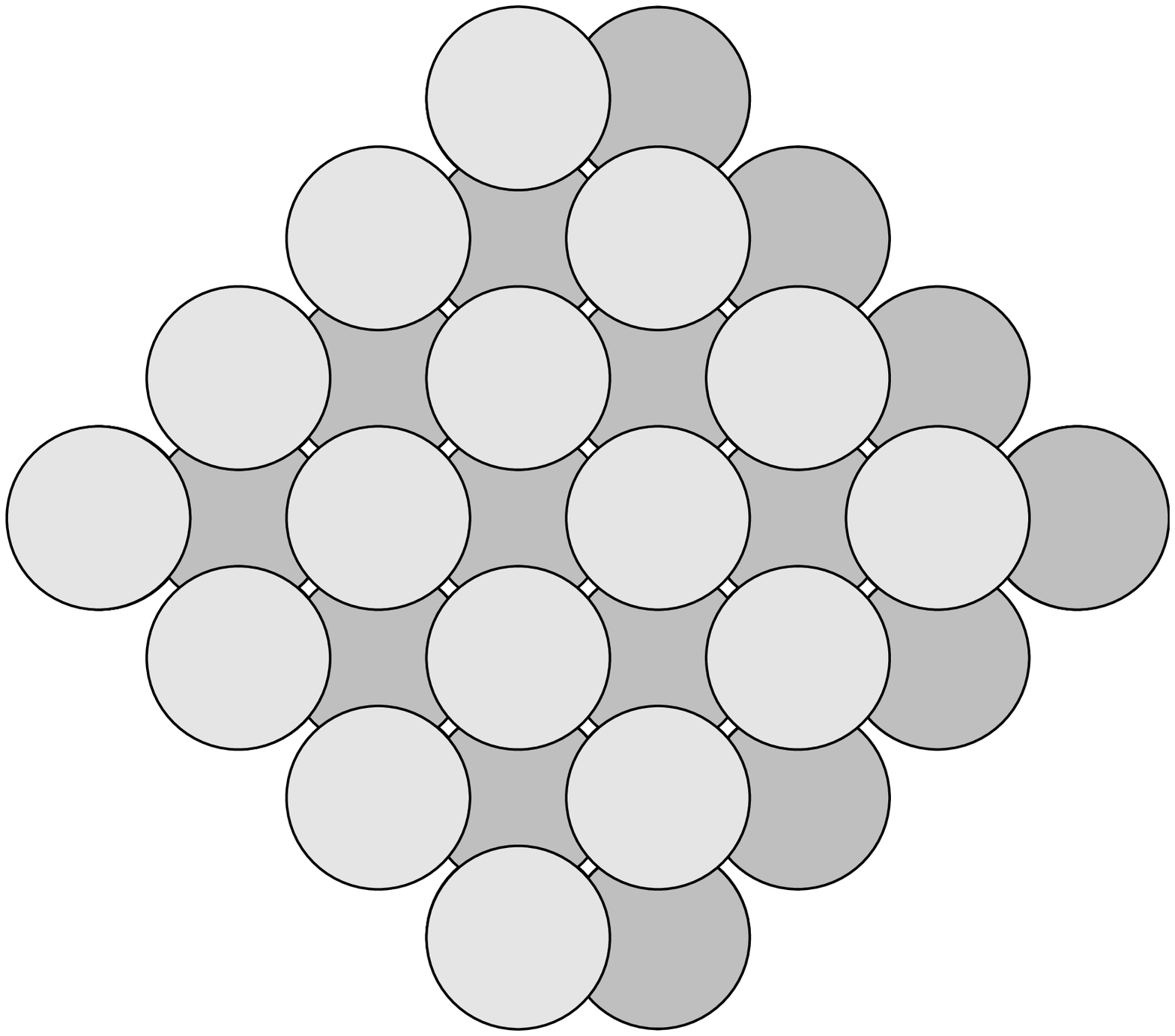,height=1.7in}}
\centerline{\psfig{figure=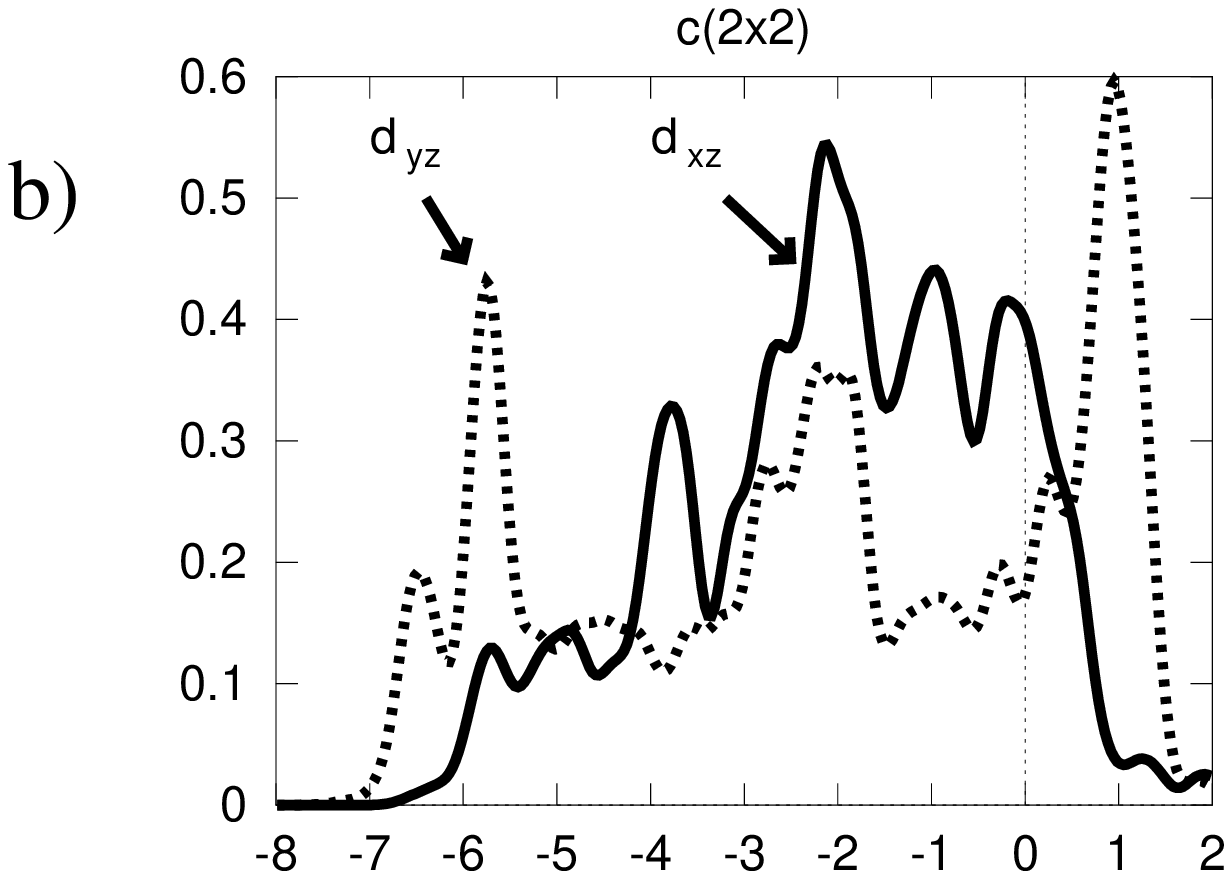,height=1.9in}
\psfig{figure=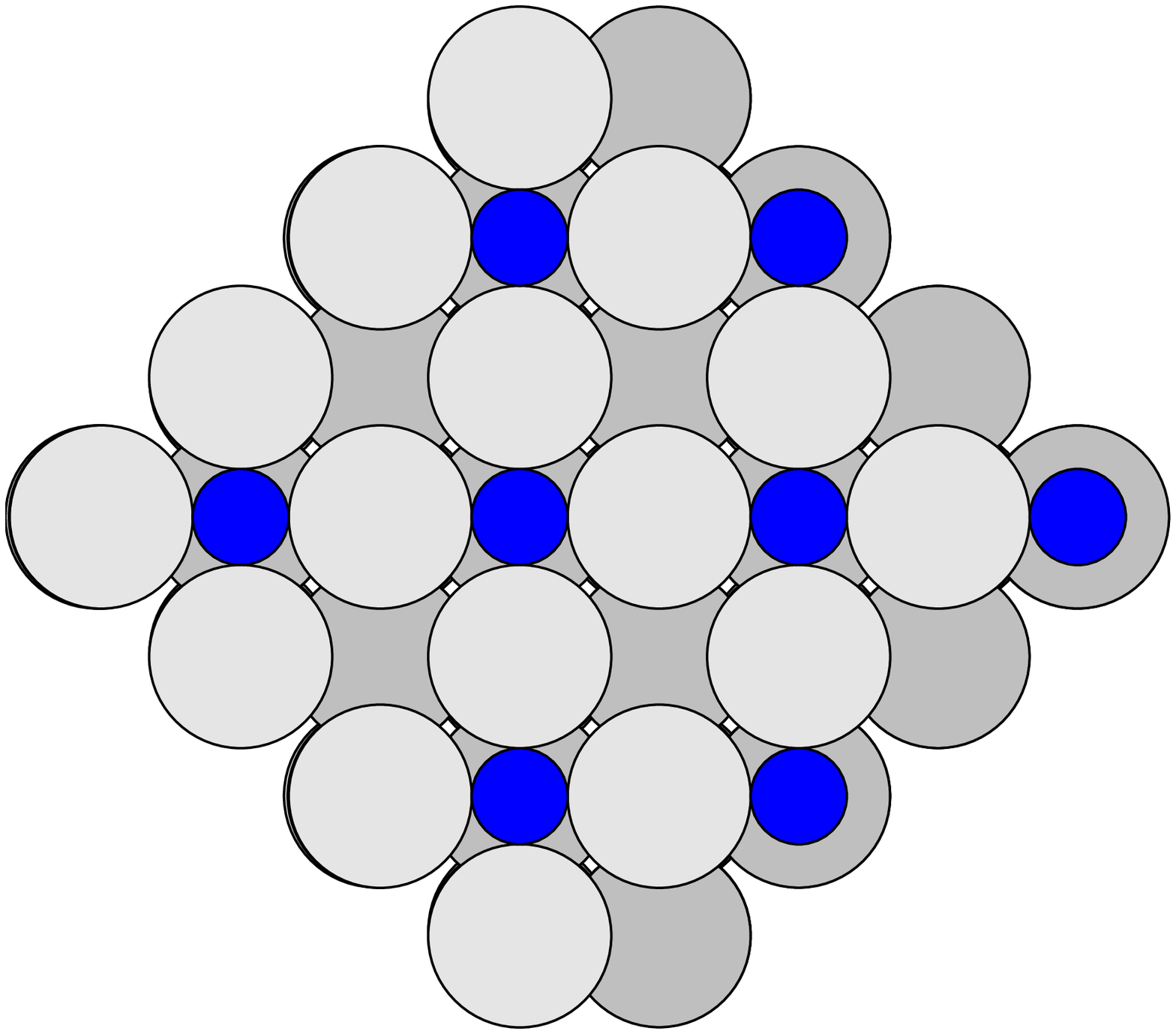,height=1.6in}}
\centerline{\psfig{figure=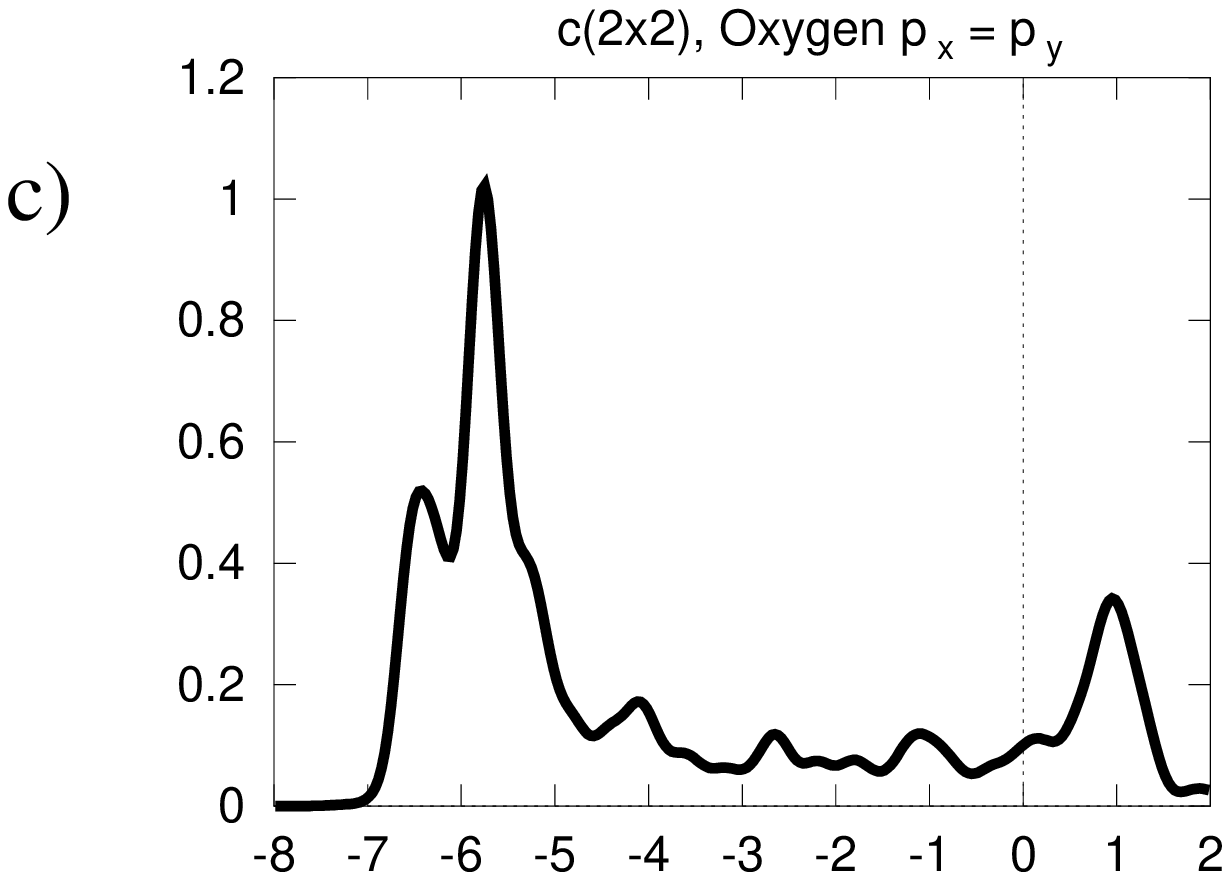,height=1.9in}
\phantom{\psfig{figure=c2x2_1.ps,height=1.6in}}}
\centerline{\psfig{figure=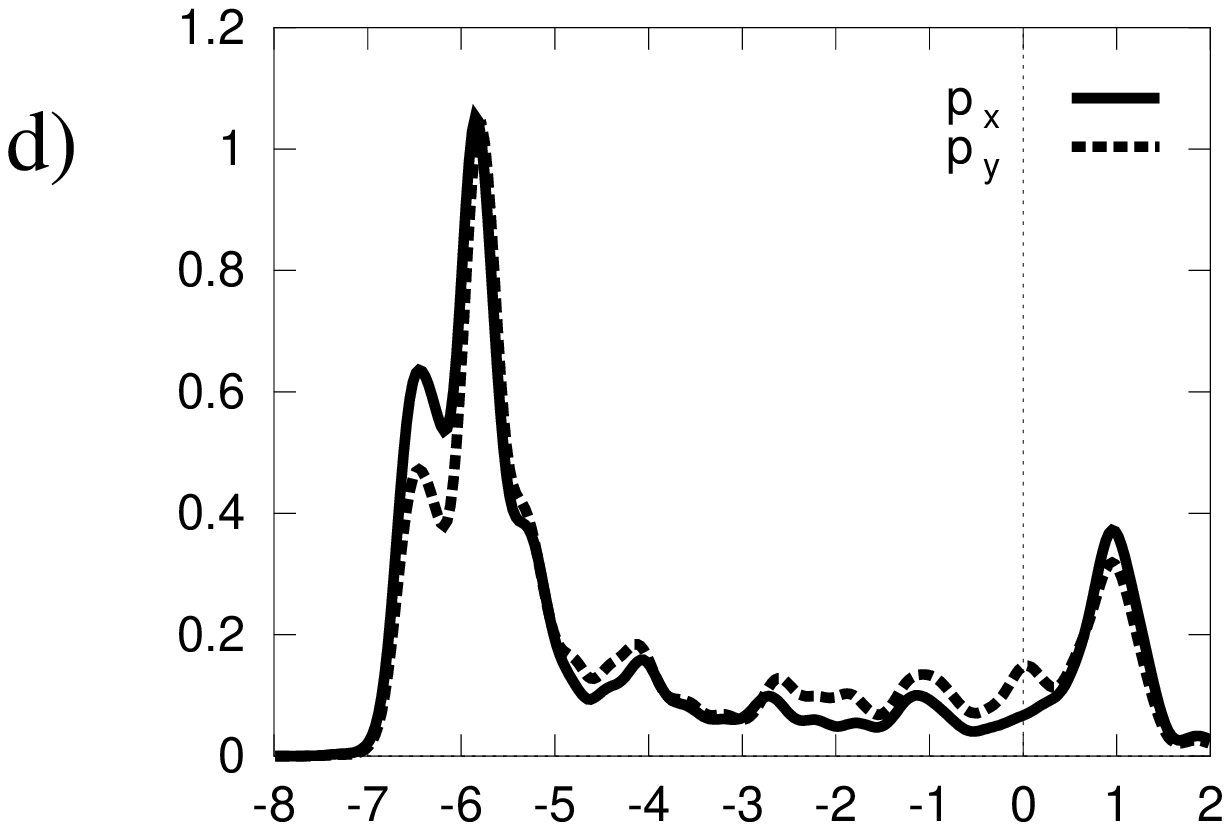,height=1.9in}
\psfig{figure=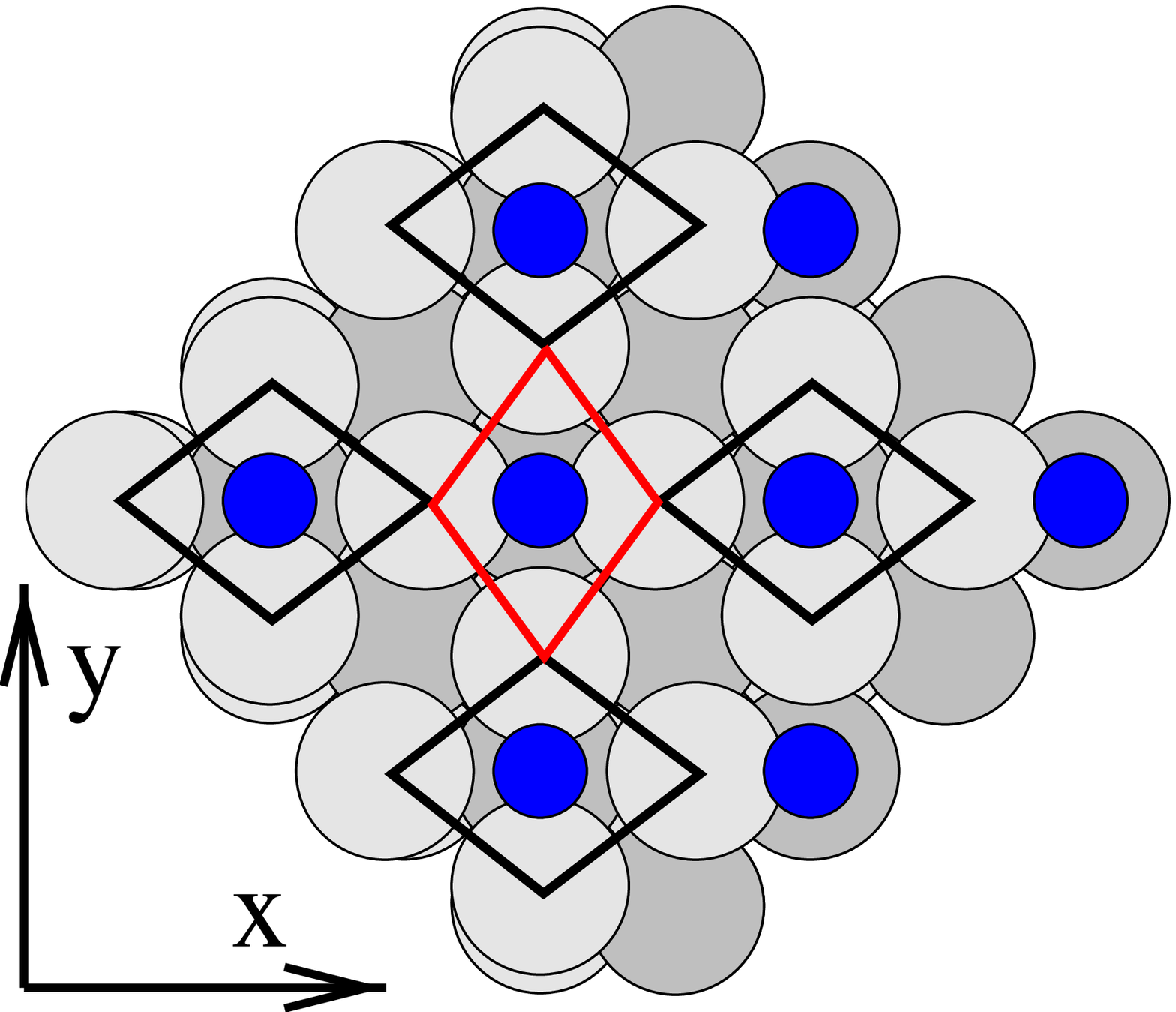,height=1.6in}}
\vskip 8pt
\caption{Local density of states projected onto surface atomic
orbitals. (a): Rh $d_{xz}$ and $d_{yz}$ orbitals of the clean
surface. (b): Same as (a), but after the deposition of half a
mono-layer of oxygen, without allowing the substrate to
reconstruct. (c) Oxygen $p_x$ and $p_y$ orbitals. (d) Same as (c), but
after the clock reconstruction has taken place.
}\label{ldosrho}
\end{figure}

The LDOS is the projection of the density of states onto localized
atomic orbitals. In Fig. \ref{ldosrho}a we display the Rh$(4d_{xz})$
LDOS of one rhodium surface atom resulting from a calculation with the
clean surface; this is degenerate with the Rh$(4d_{yz})$ orbital
because of symmetry. The energy is referred to the Fermi level. In the
panel just below (b) we display the same two orbitals after half a
monolayer of oxygen is deposited, without allowing the substrate to
reconstruct. The Rh$(4d_{xz})$ orbital points towards the square site
left empty, and its LDOS is very similar to what it would be for the
clean surface. The Rh$(4d_{yz})$ orbital is directed towards the
oxygen adsorption site, and the shape of its LDOS changes completely
upon adsorption: the most important feature is the presence of two
peaks, one at $\approx -6$ eV and the second above the Fermi energy,
at $\approx 1$ eV. In the (c) panel we display the LDOS projected onto
O$(2p_x)$ orbitals, which are degenerate with the O$(2p_y)$ orbitals
because of the symmetry of the oxygen site. This LDOS is characterized
by a bonding and an anti-bonding peak, at energies of $\approx -6$ eV
and $\approx +1$ eV respectively, which are in the same positions of
the peaks observed in the Rh$(4d_{yz})$ LDOS. We interpret this fact
as the evidence of the presence of a strong covalent bond between the
oxygen atoms and the rhodium surface atoms. The different weight
between the bonding and the anti-bonding peaks also indicates that
there is a charge transfer from the substrate to the oxygen, so that
the bond is partially ionic. This fact is also supported by work
function calculations, which show an increase of $\approx 0.6$ eV when
the surface is covered by oxygen. This imply that an excess of
negative charge is present on the surface. In the last panel of the
figure we display the LDOS projected onto the O$(2p_x)$ and O$(2p_y)$
orbitals for the reconstructed surface. The reconstruction lifts the
$xy$ symmetry, and the O$(2p_x)$ orbital---which corresponds to the
shorter O-Rh bond---becomes more occupied than the O$(2p_y)$
orbital. We conclude that the mechanism of the reconstruction is due
to the re-bonding of the oxygen with the surface atoms, which tend to
shorten the O-Rh bond. This shortening could also be realized by a
penetration of the oxygen deeper into the site, but this does not
happen. The reason is maybe the excess of negative charge on the
oxygen atoms, which prevents their penetration into the electronic sea
of the metallic surface.

While the present work was being completed, we learnt that our
theoretical predictions have been recently confirmed by LEED I-V
experiments \cite{ELETTRA}. In particular, the analysis of the LEED
I-V spectra indicated that the reconstruction of the Rh surface upon
oxygen adsorption is indeed of the {\it white} type. Furthermore, the
stable adsorption site of the oxygen adatoms was found to be
asymmetric with respect to the center of the rhombus, with a
displacement $\delta_O = 0.29 \pm 0.15~\rm\AA$, to be compared with
our theoretical prediction, $\delta_O=0.35~\rm\AA$. These data are in
partial agreement with the recently published results of an experiment
based on a combination of LEED with low-energy alkali ion scattering
and recoil spectroscopy \cite{shen}. In Ref. \cite{shen} no evidence
was found of the {\it asymmetric white} clock reconstruction predicted
by us and confirmed in Ref. \cite{ELETTRA}. Furthermore, the height of
the O adsorption site off the metal surface was found to be smaller
than our predictions which, however, are again in agreement with the
findings of Ref. \cite{ELETTRA}.

At variance with oxygen, nitrogen does not induce any reconstruction
upon adsorption on Rh(001). The equilibrium distance of the
adatom from the metal surface is rather large, as it is the case for
oxygen (see Table \ref{tab:riassunto}). For carbon we have found a
metastable state with a surface reconstruction of the {\it black} type
(Fig. \ref{ricostruzione}a) and a rather small distance between the
ad-atoms and the metal surfaces (see the discussion of the {\it black}
reconstructions typical of Ni in the next subsession). However, the
energy of this local minimum is higher than that of the
unreconstructed surface, which corresponds to a larger ad-atom/surface
equilibrium distance. The presence of this second stable deep site,
even though the energy is higher, suggests that the choice of the
actual adsorption site is a tradeoff between the chemical energy gained
due to the larger number of ad-atom-Rh bonds in the deep site, and the
elastic energy lost due the distortion of the surface when this site
is occupied. We shall comment more on this issue in the next section.

\subsection{Nickel}

In agreement with experimental data \cite{oed}, for O/Ni(001) we did
not find any reconstruction at all. In this case the only effect that
has been observed is a buckling in the second metal layer, whose atoms
are not all coplanar but shifted up or down according to whether the
fourfold site lying above is empty or filled
\cite{oed,oed1,oed2,oed3,kopatzki}. This behavior is consistent with
our interpretation of the O/Rh(001) reconstruction: the nickel lattice
parameter is quite smaller than that of rhodium, and therefore the
adsorption site is already small enough for the oxygen ad-atoms to
bind with neighboring metal atoms without inducing any reconstruction.
For the carbon- and nitrogen-covered surface we have found a
reconstruction where---contrary to O:Rh(001) and in agreement with
experiments findings \cite{onuferko79,wenzel}---it is the {\it filled}
squares which rotate, i.e. a {\it black} reconstruction
(Fig. \ref{ricostruzione}a). In the case of carbon, we find that no
energy barrier exists between the unreconstructed and the
reconstructed structures, whereas such a barrier has been detected for
N:Ni(001)---though its value has not been determined.

Going back to the results summarized in Table \ref{tab:riassunto}, an
inspection of the results for $d_{01}$ (the distance between the
adsorbate atoms and the first metal layer) is particularly
instructive. In those cases where a {\it black} reconstruction occurs
(C:Ni and N:Ni), $d_{01}$ is small ($\approx 0.1~\rm\AA$), thus
indicating that the adsorbates penetrate the first metal layer. The
penetration of the adsorbate into the metal determines an {\it
outward} (with respect to the center of the square) force acting on
the metal atoms at the corners of the {\it black} squares. This can be
alternatively be described as an {\it inward} force as seen from the
center of the {\it white} squares. In the case of a {\it white}
reconstruction (O:Rh), instead, the ad-atom stays out of the surface
($d_{01} \approx 1~\rm\AA$), and they pull directly the neighboring
atoms towards the adsorption site (rather than pushing them away from
it) in order to strengthen the O-metal bonds. In both cases, the
square which is `squeezed' by the adsorption process ({\it black} or
{\it white}, depending on the case), displays an instability towards a
rhomboid distortion which is partially opposed by the elastic reaction
of the substrate.

\begin{figure}
\def\baselinestretch{1}
\centerline{\psfig{figure=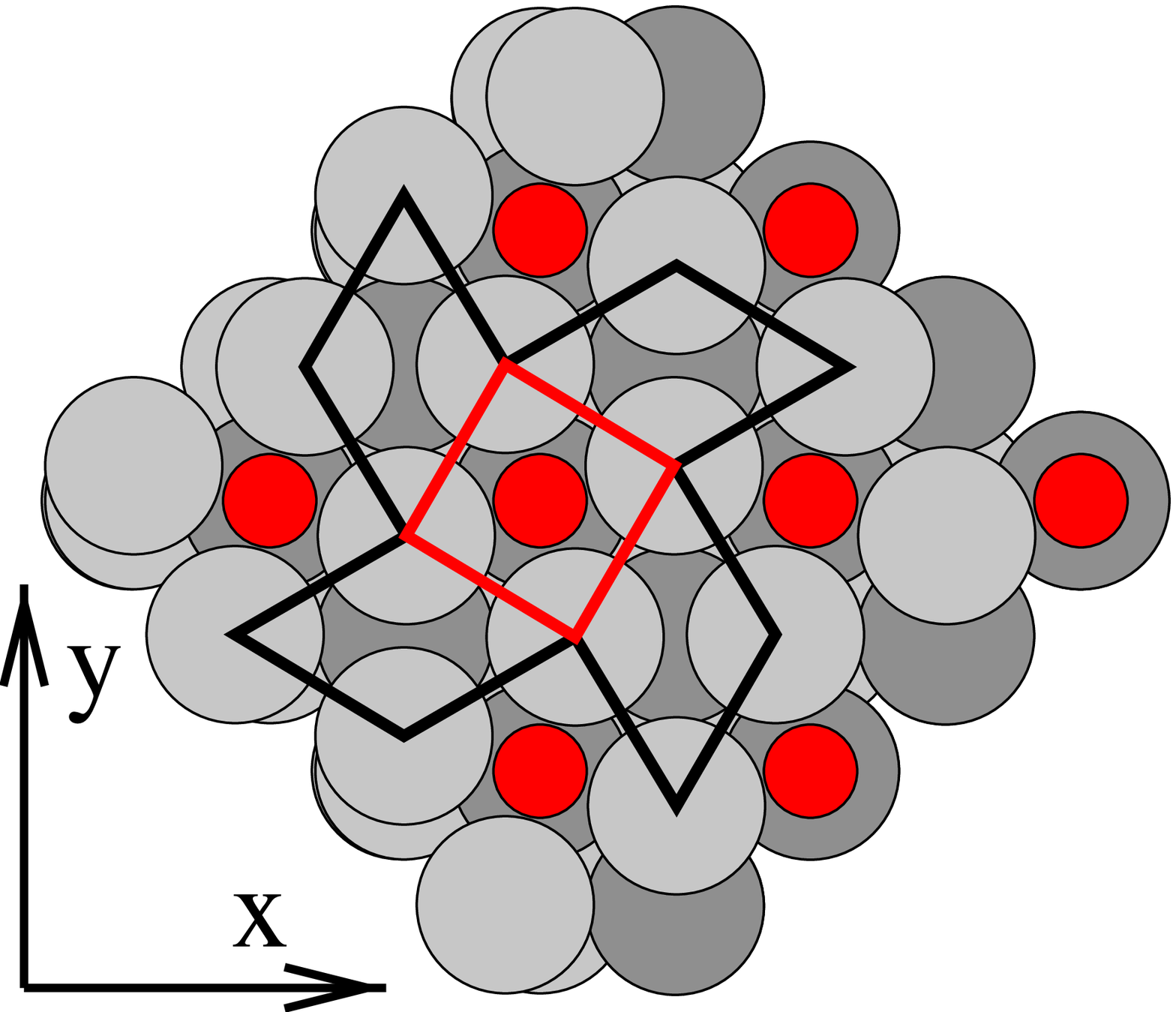,height=1.8in}}
\centerline{\psfig{figure=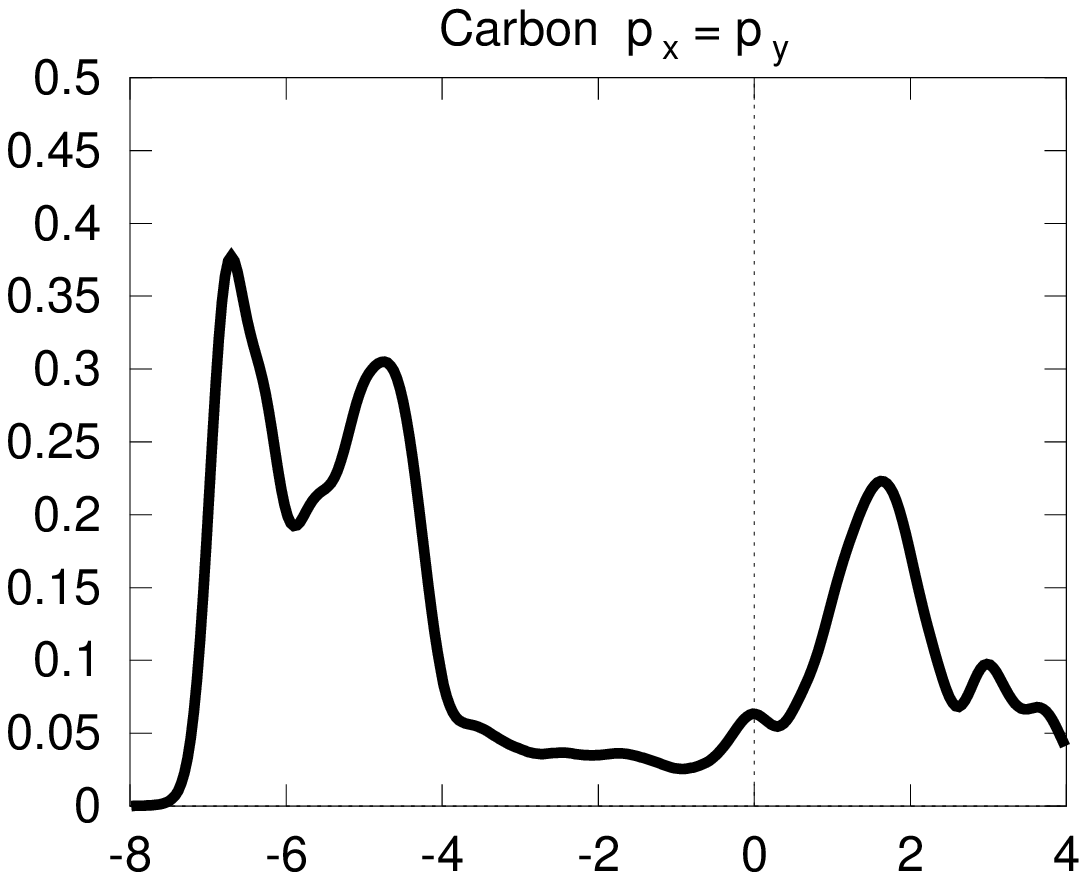,height=2.1in}
\psfig{figure=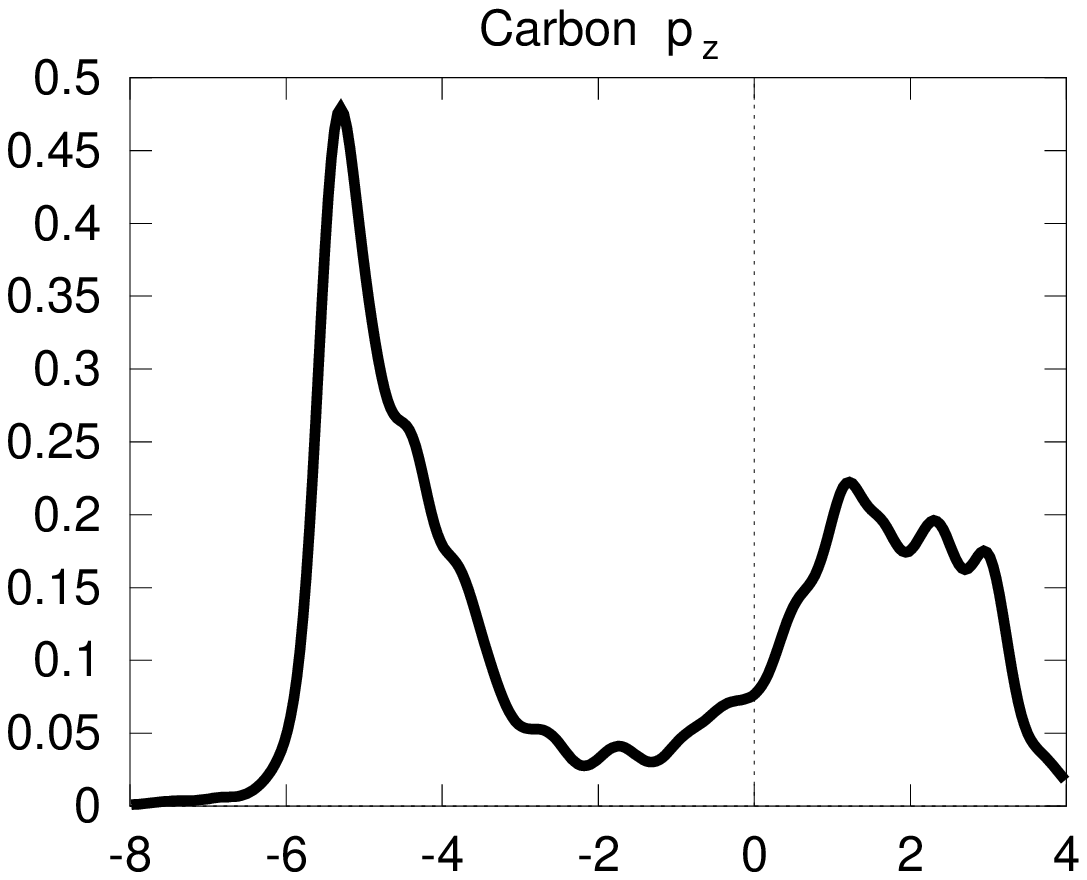,height=2.1in} \hskip 20pt}
\centerline{\psfig{figure=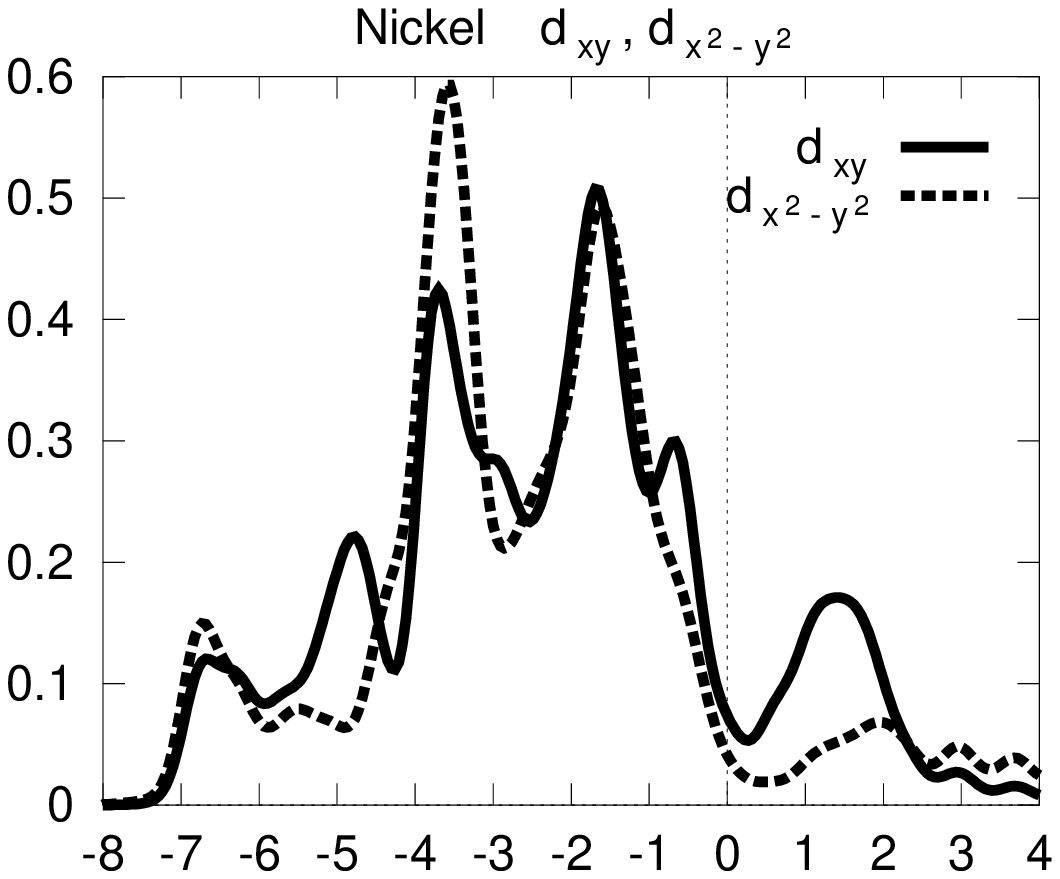,height=2.1in}
\psfig{figure=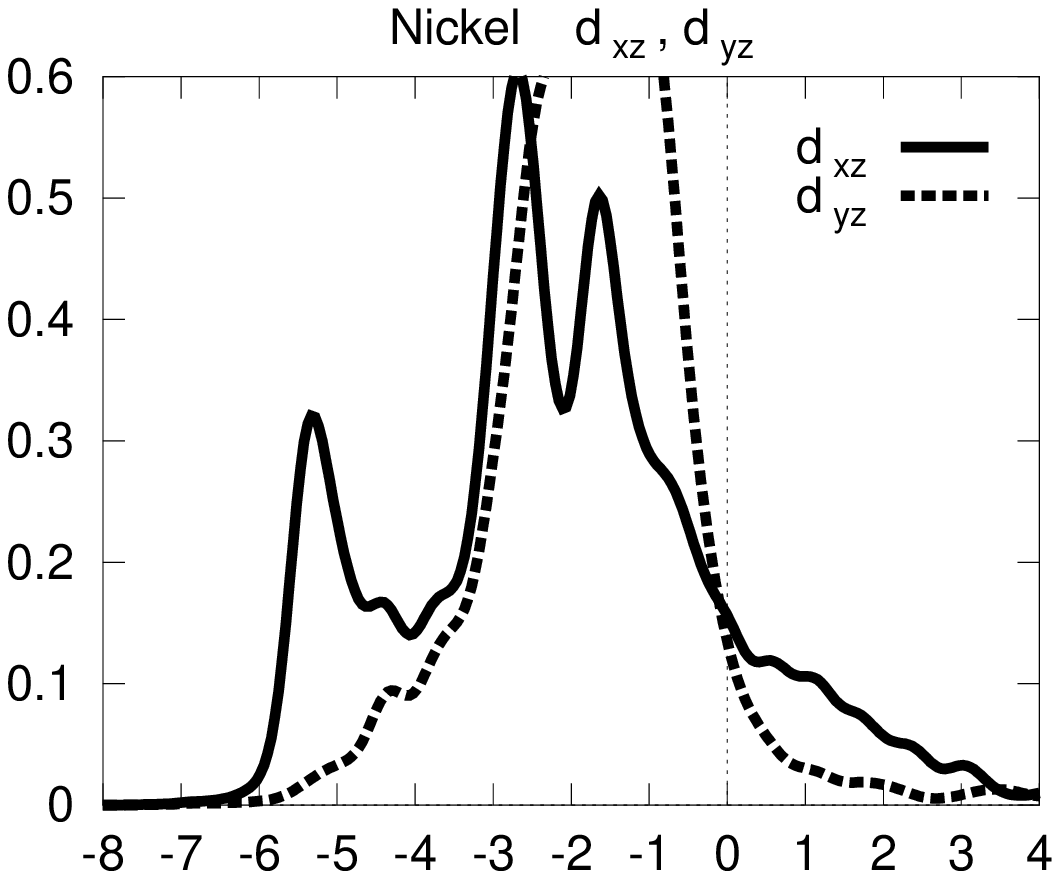,height=2.1in} \hskip 20pt}
\centerline{\psfig{figure=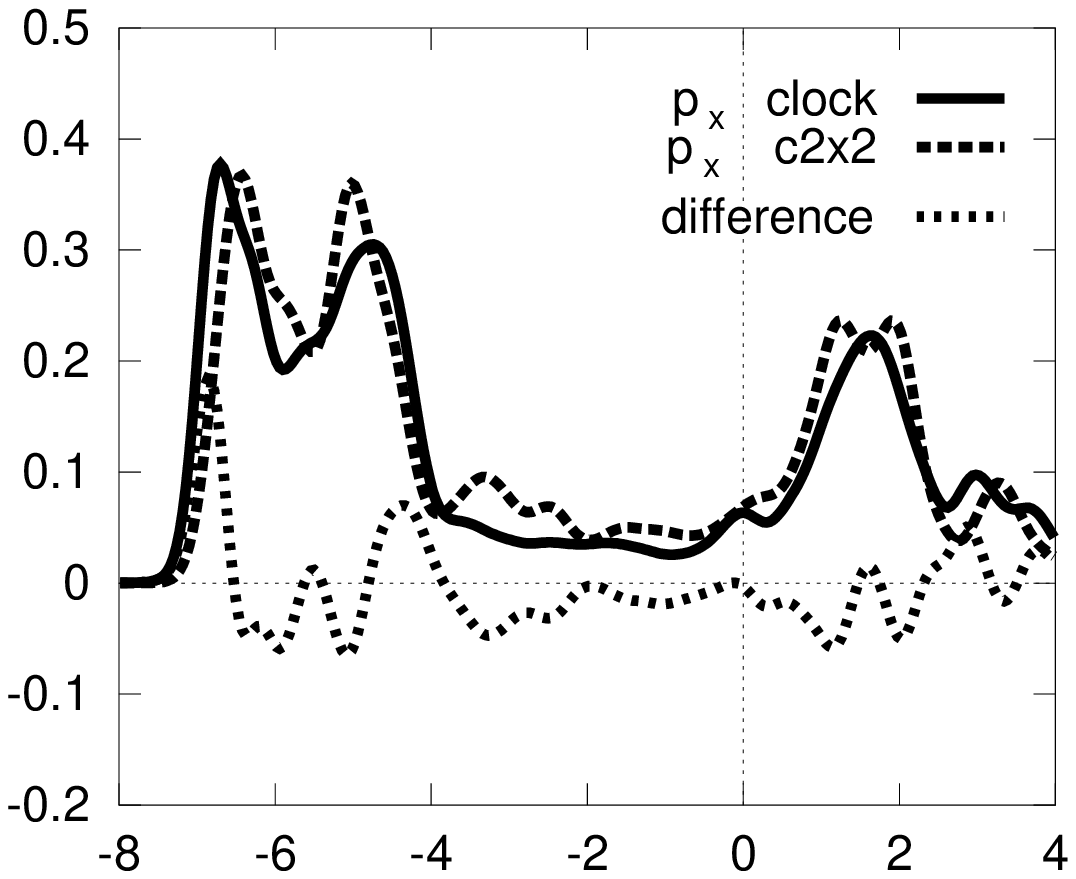,height=2.1in}
\psfig{figure=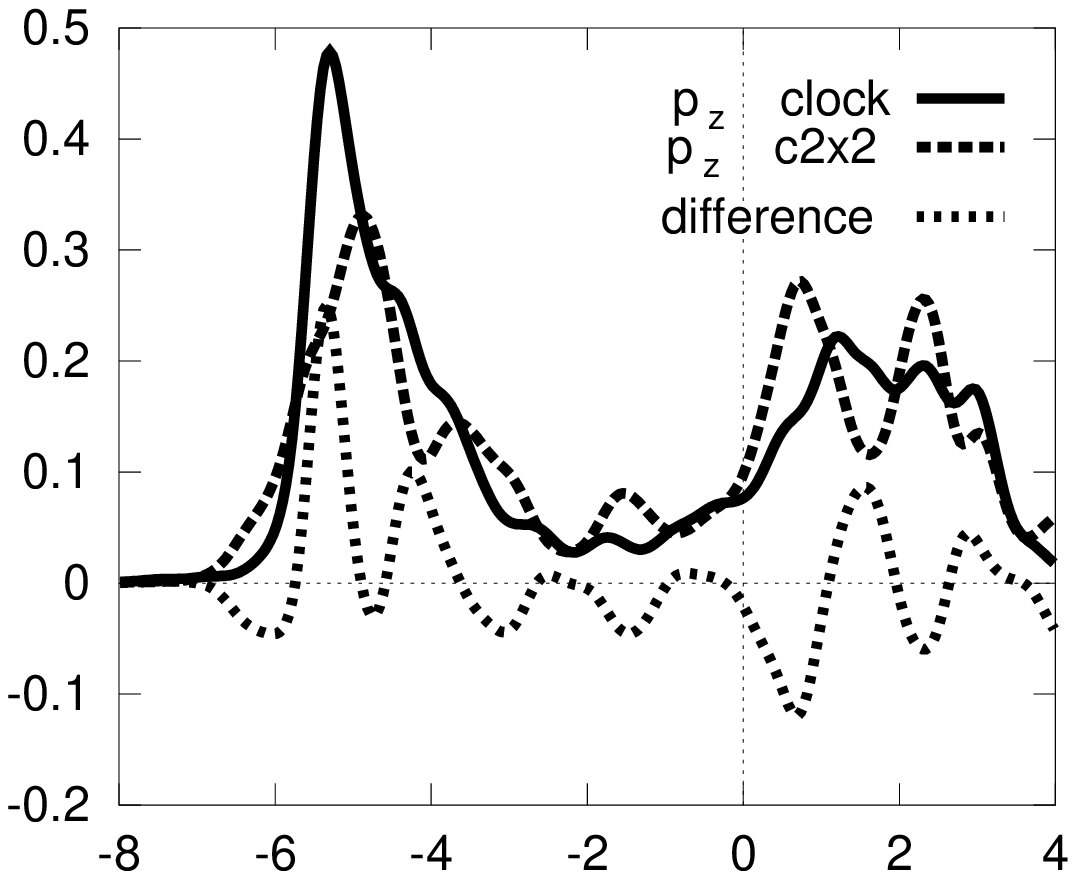,height=2.1in} \hskip 20pt}
\vskip 8pt
\caption{Upper panels: local density of states of the C/Ni(001)
surface, projected onto C $p_x$ or $p_y$ (left) and C $p_z$ (right)
atomic orbitals. Middle panels: same as above, but projected upon Ni
$d_{xy}$ and $d_{x^2-y^2}$ orbitals (left) and Ni $d_{xz}$ and
$d_{yz}$ orbitals. Lower panels: difference between the LDOS projected
upon C $p$ orbitals before and after the clock reconstruction takes
place.}\label{proj}
\end{figure}

It is worth noting that in the cases where a {\it black}
reconstruction occurs---and, hence, $d_{01}$ is small---the
unreconstructed structure is a metastable (N/Ni) or unstable (C/Ni)
equilibrium structure characterized by a rather large value of
$d_{01}$ (equal to 0.48 \AA\ for both N/Ni and C/Ni). In order to
understand the chemistry of the bond between carbon and nickel we have
analyzed the local density of states for the unreconstructed and
reconstructed surfaces, which we display in Fig. \ref{proj}. In the
two upper panels we show the LDOS projected onto the C $p_x$ or $p_y$
orbitals which are degenerate by symmetry (left) and onto the C $p_z$
orbital (right). In both cases a bonding and an anti-bonding structure
can be clearly identified. In the case of the C $p_{x,y}$ orbitals,
the bonding structure is resonant with similar structures observed in
the LDOS projected onto the Ni $d_{xy}$ and $d_{x^2-y^2}$ orbitals,
while the C $p_z$ bonding structure is resonant with Ni $d_{xz}$
(middle panels). Inspection of the difference between the C
$p_{x,y}$-- and C $p_z$--projected LDOS before and after
reconstruction shows that the gain in binding energy upon
reconstruction mainly comes from a shift of the C $p_z$ LDOS towards
lower energies, thus indicating the establishment of a covalent-like
bond between the C $p_z$ and Ni $d_{xz}$ orbitals.

\section{Discussion and Conclusions}\label{discussion}

The behavior of the two surfaces studied here (Ni(001) and Rh(001)) is
apparently very similar---they have the same STM images and LEED
patterns---but a more accurate inspection of their structural and
electronic properties reveals many interesting differences. The most
important one is that, in spite of the fact that the adsorption site
of the adatoms is the same, the kind of reconstruction induced by them
is different for O/Rh(001) and for C/Ni(001) or N/Ni(001). In the
first case the oxygen site is deformed into a rhombus ({\it white}
reconstruction), while in the second case the adsorption sites remain
square, and it is the empty sites which are deformed into rhombi ({\it
black} reconstruction, see Fig. \ref{ricostruzione}a). Secondly, on
rhodium the distance of the oxygen atoms from the surface is not
appreciably different for the unreconstructed and the reconstructed
surfaces, while on nickel we have found that when the reconstruction
takes place carbon and nitrogen atoms become almost coplanar with the
substrate ones, thus becoming essentially five-fold coordinated. The
carbon atoms penetrate into the nickel substrate even at zero
temperature, and they arrange themselves so as to be almost coplanar
with the first nickel layer; the nitrogen atoms behave similarly, but
a barrier is found to exist which gives rise to a meta-stable
equilibrium structure in which the ad-atoms do not penetrate and the
surface does not reconstruct. These facts indicate that the mechanism
of the reconstruction is related with the atomic penetration into the
first surface layer: the sites which accommodate the ad-atoms tend to
enlarge, and this results necessarily in a rhomboid distortion of the
empty sites. This distortion has a cost in term of elastic energy. We
argue that the surface reconstructs when the chemical effects are
larger than the elastic effects. Upon C or N adsorption, the Ni(001)
surface reconstructs, while Rh(001) does not. The reason is probably
the different stiffness of the two metals: rhodium has larger elastic
constants and thus it does not reconstruct. In the case of the oxygen
the mechanism of the reconstruction is completely different, being due
to the O--Rh re-bonding, and results in a different reconstruction
pattern.

Further details on this work can be found in the PhD thesis of one of
us (DA) \cite{PhD}.

\acknowledgements We are grateful to Renzo Rosei for inspiring this
work and for making the results of Ref. \cite{ELETTRA} available to us
prior to publication. Many useful discussions with A. Baraldi,
G. Comelli, and R. Rosei are gratefully acknowledged. Our calculations
were performed on the SISSA IBM-SP2 and CINECA-INFM Cray-T3D/E
parallel machines in Trieste and Bologna respectively, using the
parallel version of the {\tt PWSCF} code. Access to the Cray machines
has been granted within the {\it Iniziativa Trasversale Calcolo
Parallelo} of the INFM. Finally, we acknowledge support from the MURST
within the initiative {\it Progetti di ricerca di rilevante interesse
nazionale}.

\end{document}